\documentclass[12pt,a4paper]{article}

\usepackage[T2A]{fontenc}
\usepackage[utf8]{inputenc}

\usepackage{color,amsmath,amsfonts,amssymb,amsthm,upref,mathrsfs,bm}
\usepackage{graphicx}
\usepackage{caption}
\usepackage{subcaption}
\usepackage{slashed}
\usepackage[notref,notcite]{showkeys}
\usepackage{epsfig}
\usepackage[english]{babel}
\usepackage{color}

\def\({\left(}
\def\){\right)}
\def\[{\left[}
\def\]{\right]}
\def\be{\begin{equation}}
\def\ee{\end{equation}}
\def\a{\alpha}
\def\b{\beta}
\def\g{\gamma}
\def\e{\varepsilon}
\def\k{\varkappa}

\def\o{\omega}
\def\om{\omega}
\def\O{\Omega}
\def\K{{\cal K}}

\def\bk{{\bf k}}
\def\br{{\bf r}}
\def\bv{{\bf v}}

\def\bp{{\bf p}}

\def\no{\varpi}
\def\nk{\chi}
\def\bnk{{\bm \chi}}

\def\G{{\bm \Gamma}}
\def\Im{\mathop{\rm Im}}
\def\Re{\mathop{\rm Re}}
\def\exp#1{\mathop{\rm exp}\nolimits\hskip-.7mm\left\{#1\right\}}

\def\tr{\mathop{\rm tr}}

\def\Ref#1{(\ref{#1})}
\def\foo{\footnote}

\begin{document}

\title{On  astigmatic solutions of the wave and the Klein-Gordon-Fock
equations with exponential fall--off}

\author{I. V. Fialkovsky$^\dag$\footnote{ifialk@gmail.com}, M. V. Perel$^\ddag$ and A. B. Plachenov$^\|$
\\{$^\dag$ CMCC-Universidade Federal do ABC, Santo Andr\'e, S.P., Brazil}
\\ $^\ddag$ Department of Mathematical Physics, Physics Faculty,\\
Saint-Petersburg State University, Russia\\ and Ioffe Physical-Technical Institute, Russia
\\
{ $^\|$ Moscow State Technical University of Radioengineering,}\\{Electronics and Automation, Russia,}\\{ and Saint-Petersburg State University of Aerospace Instrumentation, Russia}}
\maketitle

\begin{abstract}
Highly localized explicit solutions to  multidimensional wave and
Klein--Gordon--Fock equations are presented. Their Fourier transform is also found explicitly.
Solutions depend  on
a set of parameters, and demonstrate astigmatic properties. Asymptotic analysis for large
and moderate time shows that constructed solutions have Gaussian localisation near a point moving with the group speed.
\end{abstract}

\section{Introduction}

Seeking localized solutions to (non-) linear differential
equations has a very long history, starting probably with
the famous observation by John Scott Russell of a solitary
wave in the Union Canal \cite{Russell}.
 Since then, many a research were made. An interest to the
 theoretical study of localized solutions of linear equations
 was renewed after the discovery of lasers and   further
 progress in  technologies of emitting ultra-short pulses. Nowadays, there are also
numerous  potential  applications of such solutions, for
example,
for the localized 
low-loss energy transmission, communication, medical
imaging or
remote sensing.  The state of the art in this field, along with its history, is presented in recent books \cite{Recami08,Recami13}.

In free space (or non-dispersive 
media) the problems of wave propagation are usually
described using
the wave equation  (WE)
 \be \label{wave-eq}
    \square \phi \equiv \partial^2_{t}\phi  - \Delta \phi =0,
\ee
where $ \partial_{t}(\cdot)\equiv \frac\partial{\partial t}
(\cdot)$, etc., and
$$
    \Delta \phi \equiv \partial^2_{x}\phi +
    \partial^2_{y}\phi  +  \partial^2_{z}\phi
$$
is three dimensional Laplace operator.  Here and throughout
the paper we put the constant of speed of light equal to
unity, $c=1$.

On the other hand, the waves in  dispersive  medium are
often described by the Klein--Gordon--Fock equation (KGFE)
 \be \label{KGFE}
    (\square+m^2) u  =0,
 \ee
where $m$ is a mass parameter having different physical
meaning in different systems. The KGFE is important in
studying of electromagnetic waves in the isotropic cold
collisionless plasma \cite{Ahiezer},   waves of charge
density in Drude metals, or the high-frequency acoustic
waves in the gas of charged particles
treated in the hydrodynamic approach \cite{FrankKamenetsky}.

A relatively recent spike in activity in investigating
localized solutions was promoted by the paper by
Brittingham \cite{Brittingham83}, who outlined a new
one--parametric family of beam-like localized explicit
solutions to the WE which he named the focus wave mode
(FWM, which we also call the Gaussian beam).  This solution
is localized in the Gaussian way along a straight line in
space and has infinite energy.
This solution cannot be obtained by the separation of
variables in coordinates $\br = (x,y,z)$ and time. However
 introduction of new variables
\be\label{albe}
\alpha = z -t, \quad \b = z +t,
\ee
where $z$ is  coordinate along the propagation axis,
enables one to do this.
 The KGF analogue of the Brittingham FWM for the WE was
 given by Ziolkowski in \cite{Ziolkowski85} by the
 separation of variables (\ref{albe}).
 In \cite{Ziolkowski89} he also suggested to seek other
 solutions as weighted superpositions over a free parameter
 of FWM and  in doing so he obtained a solution of   finite
 energy and a power--law localization.
Another way of construction of highly localized solutions,
the so called bidirectional representation, is based on
the Fourier integral in new variables (\ref{albe}). Taking
the Fourier weight in the proper way Besieres, Shaarawi and
Ziolkowski \cite{Besieres89} , Donelly and Ziolkowski
\cite{DonnZiol92}  found new solutions with finite energy
of  WE and KGFE respectively.

All highly localized solutions both for the WE and for
KGFE spread propagating.
There were found however the so called undispersive
solutions, which propagate without spreading,  but have
a power-law localization from their moving amplitude
maximum.
For the WE these are the  Bessel beams, solutions found  by
Durnin \cite{Durnin1987} by separation of variables
in initial space-time coordinates, and the X--waves found by
weighted superposition of Bessel beams \cite{Lu-Greenleaf}.
For the KGFE solutions  the same property is possessed by
MacKinnon's solution \cite{MacKinnon}  which can be found
by   combination of the separation of variables and the
Lorentz transformations.

The particle--like exact solution to the WE which decreases
exponentially in all directions away from a point moving
along a straight line was first presented by Kiselev and
Perel in \cite{KiselevPerel00}. Perel and Sidorenko
\cite{PerSid} considered this solution from the point of
view of wavelet analysis, investigated numerically the
uncertainty relation for this solution and found explicitly
its  Fourier transform. In
\cite{PerSid}  the solution was also analyzed  from the point of view of complex sources. It
was shown that it can  be generated by a pulse source moving
with the speed of wave propagation.  An integral
representation of this solution  in terms of Gaussian beams
due to ideas of \cite{Ziolkowski89} was given by Perel and
Fialkovsky in the paper  \cite{PeFi01} which was however
mainly   devoted  to the KGFE.
 In  \cite{PeFi01} it was also suggested a  class of
 explicit exponentially localized  
 packet-like solutions
 for KGFE and investigated some of their properties.
 One of solutions from the class obtained in
 \cite{PeFi01} coincides with one from \cite{Besieres00}.
 Kiselev, Plachenov and Chamorro-Posada  \cite{KisPlach12}
 created astigmatic  beam-like and packet-like  solutions
 of the WE.

Finally, we shall also mention the work by Overfelt
\cite{Overfelt91} who got a class of solutions which
generalizes  
Gaussian beams of \cite{Brittingham83}  and the Bessel
beams of
\cite{Durnin1987} and which  have better localization
 near the
propagation axis than Gaussian beams.
In Besieres, Shaarawi and
Ziolkowksi \cite{Besieres00} there was  suggested a new
method of
design of solutions of three dimensional KGFE reducing
them to a solution of one dimensional  KGFE equation with new
'time' and 'coordinate' containing an arbitrary function. On this way
they found the counterparts of the Gauss--Bessel pulses
\cite{Overfelt91} and some other solutions.
Intrerested reader can find more detialed refences and discussion of various constructions of the localized solutions both for the WE and the KGFE in books \cite{Recami08,Recami13} and reviews \cite{Recami09,Tachions}.

The present work is a continuation of works \cite{PeFi01} and
\cite{KisPlach12}. We present nonseparable  solutions with
exponential localization, 
both beam-like and particle-like ones. The former solutions
have
Gaussian localization near a straight line and traveling
wave-fronts. The latter ones in addition to localization
near a line
are  localized in a Gaussian manner near a point moving
along this
line.

The structure of the paper is following. In the next section
we revisit known results on exact exponentially localized
solutions for the wave equation. By doing so, we also
construct the astigmatic generalizations of the known
beam--like and particle--like solutions for the WE. In
Section \ref{KGFeqn} we apply the developed methods to
the construction of the multidimensional astigmatic
solutions to the Klein--Gordon--Fock equation. We proceed
by asymptotic investigation of the KGF solutions in Section
\ref{AsyInv_sect} where we consider both small/moderate and
large times regimes and discuss the choice of parameters
which enable us to govern the localization properties of
solutions. We conclude our research by obtaining the Fourier
transformation of all constructed families of solutions in
Section \ref{Fourier-analysis}, and by providing some final
remarks and numerical studies of the solutions in Section
\ref{Discu_sec}. In Appendix \ref{Asymptotic-KGF} we present
the asymptotic investigation of obtained solutions in the
Fourier domain. All our results are valid for the
space--time with any number of spatial dimensions.

\section{Wave Equation Revisited}\label{Wave-Equation}
In this section we construct a generalizations of the
known  localized  exact solutions to the WE
\be
    \partial_t^2 \phi - \Delta_{n} \phi =0,
    \label{wave-eqN}
\ee
 where $ \Delta_{n}$ stands for $n$ dimensional
 Laplacian operator.  By doing so we also revive the
 necessary techniques to be used also for the
 Klein--Gordon--Fock equation.
In constructing particle--like solutions of the wave
equation, we
follow the idea by Ziolkowski \cite{Ziolkowski89} and seek
such solutions in
the form of a superposition of Gaussian beams.

The exponentially localized 
solutions we are focusing on
 may be considered as  ``relatively undistorted progressive
 waves'' by Courant and Hilbert \cite{CourantHilbert62}, or
 ``nondispersive waves'' by Hillion
 \cite{Hillion92,Hillion93}.  They are of the form of a ray
 series which comprises one term only
 \be
    \phi=g(\br,t)f(\theta),
\label{rel-dist}\ee
where $\theta$ is a solution of the eikonal equation
for WE
\be
(\partial_t\theta)^2 - (\nabla \theta)^2 =0, \label{eikon}
\ee
function $g(\br,t)$ depends on the form of $\theta$ and
satisfies two equations
\be
(\partial_t\theta)(\partial_tg_0) - \langle\nabla \theta,
\nabla g_0\rangle + g_0 \square \theta = 0,
\quad (\partial_t g)^2 - (\nabla g)^2 =0, \label{transp2}
\ee
and $f(\theta)$ is an arbitrary function. Here
$\langle a b\rangle=\sum_{i=1}^n a_i b_i$.

The focus wave modes by Brittingham \cite{Brittingham83}
 belonging to the so called Bateman--Hillion class are
 based on the following  eikonal   \cite{Hillion92,Bateman55}
\be
    \theta=\a+\frac{\br_\perp^2}{\b-i\e},
    \quad \br = (\br_\perp, z),
    \label{theta}
\ee
where $z\equiv x_n$, $\alpha$ and $\b$ are defined in
(\ref{albe}),
and $\e$ is an arbitrary positive constant.
The function  $g(\br,t)$ in this case is of the form
$ g(\br,t)= {(\b-i\e)^{-d/2}}$,  $d = n-1$. Choosing the
arbitrary function $f(\theta)$ as a pure exponent
\cite{Brittingham83,Kiselev83}
$
    f(\theta) = \exp{i \eta \theta}, $
where $\eta$ is a positive parameter we obtain FWM
\cite{Brittingham83}, or Gaussian beam, which  reads
\be
    \phi\equiv\frac{e^{i\eta\theta}}{(\b-i\e)^{d/2}}
    =\frac{1}{(\b-i\e)^{d/2}}
    \exp{i\(\eta\a+\frac{\b}{\e}\frac{\br_\perp^2}{\Delta_\perp^2}\)
            -\frac{\br_\perp^2}{\Delta_\perp^2}}.
            \label{BritKis}
\ee
We separated imaginary and real parts in  the exponent
to stress
exponential localization 
of the solution near $z$-axis, with
$\Delta_\perp=\sqrt{\frac{\b^2+\e^2}{\eta\e}}$
we denote the
width of the  beam in all transverse  directions.
Such axisymmetric solutions are called the \emph{stigmatic}
ones.

One of the obvious but far reaching generalizations of
the solution \Ref{BritKis}
is achieved by formal linear transformation of the
transverse coordinates in the eikonal function
$\theta$ \Ref{theta} (see, e.g., \cite{KisPlach12})
\be    \label{Theta}
     \theta=\a + (\br_\perp, \G(\beta) \br_\perp),
\ee
where $\G$ is a $d\times d$ complex matrix depending
on $\b$,
whose properties are to be discussed later. Astigmatic
generalization of \Ref{BritKis}   can be obtained by putting
$g(\br,t)\equiv g(z+t)= \sqrt{\det{\G}(\b)}$ and is given by
\begin{equation}
    \phi_b(\a,\b,\br_\perp)
        = c_b \sqrt{\det \G(\beta)} \exp{i \eta \theta},
        \label{waveBeam}
\end{equation}
where $c_b$  and
$\eta>0$ are arbitrary constants.
The WE (\ref{wave-eqN}) is indeed resolved by
(\ref{waveBeam}) if $\G(\b)$
satisfies the Bernoulli equation
$
      \partial_\b\G=-\G^2\,
$
and therefore reads
\be
    \G(\beta) = \G_0({\bf E}+\beta \G_0)^{-1},
    \label{GammaSol}
\ee
where ${\bf E}$ is the unity $d \times d$  matrix and
$\G_0=\G(0)$ --- constant
non-degenerate one. Function
$\phi_b$ has no singularities if  $\G_0$ does not have
nonzero real eigenvalues, and it has the Gaussian
localization around the axis $z$ if  the matrix
$\Im \G(\beta)$ is positively defined.  Both conditions
are fulfilled if $\G_0$ has a positive definite imaginary
part.
Indeed, the regularity condition is fulfilled since $\Im
({\br_\perp},\G_0 {\br_\perp })>0$ for all nonzero vectors
${\br_\perp}$ (real or complex) including eigenvectors
of $\G_0$ and
hence all its eigenvalues have positive imaginary parts.
To show the localization of solution for all values of
$\b$ we note  that  once $\G_0^{-1}$
has negative definite imaginary part,
$\Im \G^{-1}(\beta) \equiv  \Im \G_0^{-1}$ is also
negative definite matrix. Then $\Im \G (\beta)$ stays
positively defined
for all $\beta$.  We remind here that positive definiteness
of $\Im\G$ and negative
definiteness of $\Im\G^{-1}$ are equivalent:
$\Im ({\br_\perp},\G
{\br_\perp }) = \Im (\G^{-1}{\bp}, {\bp })= - \Im ({\bp},
\G^{-1}{\bp })$, where ${\bp }=\G {\br_\perp }$.

The axisymmetrical {stigmatic}  beam (\ref{BritKis}) may be
obtained from (\ref{waveBeam}) if
$
    \G (\b )={{\bf E}}{(\b-z_0-i\e)^{-1}}
$
($z_0$ and $b>0$ are real constants). The width of the
Gaussian curve in the transverse direction  depends in
this case on  the propagation direction $z$  and  time. For fixed time the width has a minimum which is called a waist.

The simplest non-axisymmetrical solutions, the
\textit{aligned simple astigmatic} ones,
 correspond to a diagonal $\G$--matrix:
$
   \G_{jk}={\delta_{jk}}{(\b-z_j-i\e_j)^{-1}}
$
with $z_j$ and $\e_j>0$ being real constants,
$j,k=1,2,\ldots n$.
 When $\G(\b)$ can be diagonalized by a orthogonal
 rotation of axes 
 the solutions are called \emph{rotated simple astigmatic}.
In two--dimensional  case it is given by
\be\label{G0-matr}
   \G= {\bf U} {\bf \Lambda} {\bf U}^{-1},
   \quad {\bf U} = \left(
           \begin{array}{cc}
             \cos\Phi & - \sin\Phi  \\
             \sin\Phi & \cos \Phi
           \end{array}
         \right), \quad {\bf \Lambda} = \left(
           \begin{array}{cc}
             \frac{1}{\b - z_1 - i \e_1} & 0  \\
            0 & \frac{1}{\b - z_2 - i \e_2 }
           \end{array}
        \right).
\ee
The level surfaces of the modulus of aligned (and rotated) 
simple astigmatic solution in the transverse space are ellipsoids  (ellipses 
if the whole space is three--dimensional).  The
directions of the main axes  of ellipses, defined by the (constant) eigenvectors of $\G$, do not depend on time and the coordinate $z$. The value of the waist of the Gaussian curve as well as its position, is different for different axes.  

The case of \emph{general  astigmatism } is caractrezed by the dependence on time and $z$ of  the direction of the main axes  of the localization ellipsoid. It happens when the eigenvectors of $\G$ (which are constant due to \Ref{GammaSol}) are complex and  do not coincide with those of $\Re \G$ and $\Im \G$. The real  eigenvectors of the latter matrices do depend on $\b$.  Then the main axes of the ellipsoids (or hyperboloids) of constant phase and of modulus levels rotate with time and/or coordinate $z$. The absolute value of the total angle of rotation is equal to $\pi$  for both
of them (see \cite{ArKo69} or \cite{BykovSilichev}). There is no definition of the waist in the case of general  astigmatism.

Arnaud  and Kogelnik realised \cite{ArKo69} that in two dimensions a general astigmatic solution can be obtained by assigning a complex value to $\Phi$ in \Ref{G0-matr}. Indeed,  (\ref{waveBeam})
is still a solution of the wave equation
(\ref{wave-eqN}) in this case but the eigenvectors of $\Re \G$ and $\Im \G$ will be different.
The solution with such $\G$ is localized in the neighborhood
of the $z$--axis if \cite{ArKo69} $\e_{1,2}$ are positive and
$
\cosh^2 (2\Im \Phi) \({(z_2-z_1)^2+(\e_2-\e_1)^2}\)   <
{\((z_2-z_1)^2+(\e_2+\e_1)^2\)}$.
 We note, that
the smaller is $\Im \Phi$, the closer is the solution to a
simple astigmatic
one.

All the above discussion of the astigmatic  properties
is equally
applicable to the particle--like solutions 
of the WE of
the next section, and to all solutions of the KGFE
considered in Section \ref{KGFeqn}.



\subsection{Particle--like solutions for the wave equation}
\label{Part_WE}
We  seek the particle-like solutions of Eq. (\ref{wave-eqN})
in the form of a superposition of Gaussian beams
$\phi_b$ obtained in  \Ref{waveBeam}
\be
\phi^{(\nu)}_p (\a,\b,\br_\perp)
    = \int_0^\infty d\eta \,  F^{(\nu)}(\eta) \,
    \phi_b(\a,\b,\br_\perp, \eta),
    \label{FnuWave}
\ee
where  $F^{(\nu)}(\eta)$ is a particular function depending
on the parameter $\nu$. We put%
\be
    F^{(\nu)}(\eta) = \eta^{-\nu-1} e^{-\g(\eta+\k^2/\eta)},
    \label{Fnu}
\ee
where $\nu$, $\k$, and $\g$ are arbitrary constants,
$\k > 0$,
$\g > 0$. Such particular choice of the spectral
function $F$ is motivated by consideration of the Fourier
transformation of one of the previously known solution to
the WE, given below in (\ref{wavePaExam}).
For the first time this spectral weight appears in
\cite{Besieres89} in regard of developing the so called
`bidirectional' representation 
for the solutions of the WE.

It can easily be shown that (\ref{FnuWave}) is
reduced to an integral representation of the MacDonald
function $K_\nu$ \cite{Gradshtein} (also called the modified
Bessel function of the second kind)
 \be\label{tabl-in}
 \int\limits_0^{\infty} x^{l-1} \exp{-\frac{a}{x} -b x} \,dx
 = 2 \( \frac{a}{b}\)^{l/2} K_l\(2 \sqrt{a b}\),
 \ee
 which is valid if $\Re{a}>0$, $\Re b>0$.
 We put $l = - \nu$, $a = \k^2 \g,$ $b=\g - i \theta$
 and obtain from \Ref{FnuWave}
\be
\phi^{(\nu)}_p (x, y, z, t)
    = c_p  \sqrt{\det \G(\beta)}\, s^\nu K_\nu(s),\qquad
    s= 2\k\g \(1 - {i\theta}/{\g}\)^{1/2} ,
    \label{wavePa}
\ee
where $\theta$ is    given by (\ref{Theta}),  and $c_p$ can
be  expressed through   numerical parameters of
(\ref{waveBeam}) as
$
    c_p = 2 (2 \k^2 \g)^{-\nu} c_b ,
$
but can also be treated as an arbitrary numerical constant.
It
is worth noting that $s$ satisfies the eikonal equation
(\ref{eikon})
 for wave equation (\ref{wave-eqN}).

For particular case of  $d=2$  and diagonal $\G_0 =
i\e^{-1}{\rm diag}(1,1)$, the formula
(\ref{wavePa}) yields a family of axisymmetric solutions
first
presented in \cite{PeFi01}. For $\nu = 1/2$ it gives
\be
\phi^{(1/2)}_p (x, y, z, t)
    = {\exp{-2\k\g\sqrt{1 - i\theta/\g}}}{(\b-i\e)^{-1}},
    \label{wavePaExam}
\ee
which was first obtained in \cite{KiselevPerel00}.
It corresponds to a following choice of $f(\theta)$ in \Ref{rel-dist}:
$
    f(\theta)=e^{-p\sqrt{1-{i\theta}/{\gamma}}}
$, 
were $p$ is a positive real constant.

All of the solutions \Ref{wavePa} have a Gaussian
localization  in
vicinity of a point running with wave velocity along a
straight line provided
the free parameters satisfy for certain relations, similar
to those
obtained in \cite{PeFi01}. 
However their  detailed
investigation 
is out of scope of the current paper.

Finally we note, that all the solutions of the form of
(\ref{wavePa}) can be interpreted as part of `arbitrary
waveform solutions' (\ref{rel-dist}) with
$f(\theta)= s^\nu K_\nu(s)$ and $g=\sqrt{\det
\G(\beta)}$. Apart from being more constructive the integral
representation (\ref{FnuWave}) used here with the weight
function (\ref{Fnu}) will be  important in construction of
the solution for the KGFE, where no waveform freedom is
available.




\section{Klein--Gordon-Fock equation}\label{KGFeqn}

Operating only in the space--time domain we construct now
both beam--like and
particle--like solutions of the multidimensional  Klein--Gordon-Fock
equation
\be
    (\partial_t^2-\Delta_n+m^2)u=0
    \label{KGFn}
\ee
where $\Delta_n$ is the Laplace operator in $n$ spatial
dimensions.


\subsection{Gaussian beams}
\label{beamsKGF}

To elaborate a  particle--like solution $u_p$ of the
Klein--Gordon--Fock 
equation as a superposition of the beam--like
solutions
$u_b$, we shall first construct the  latter ones. In doing
so, we
shall consider the solution (\ref{waveBeam}) of the WE in
spacial
dimension increased by one,
$\br_\perp=(x,y,\ldots)\to\br_\perp^\zeta=
(\zeta,x,y,\ldots)$, and
calculate its Fourier transform with respect to $\zeta $
\be
    u_b(t, z, \br_\perp ) \equiv
    \int_{-\infty}^\infty d\zeta\,
    \phi_b(t , z, \br_\perp, \zeta)\, e^{-i m\zeta},
    \label{fourier}
\ee
where $\phi_b$ 
is given by (\ref{waveBeam}) in $d+1$
dimensions. We shall further assume  that the enlarged
$(d+1)\times
(d+1)$ matrix $\tilde\G$  is such that
\be
(\br_\perp^\zeta,\tilde\G(\beta)\br_\perp^\zeta)
    =(\br_\perp ,\G(\beta) \br_\perp)+\zeta^{2}/(\b-i\e_m)
    \label{zeta}
\ee
with a positive constant $\e_m$, and $\G$ is a
$d \times d$--matrix as before. Then
$$
    u_b
    =c_b \sqrt{\det \G(\b)}\,e^{i\eta\theta} \,\,
    (\b-i\e_m)^{-1/2}  \int_{-\infty}^\infty d\zeta\,
        e^{i {\eta \zeta^2}/({\b-i\e_m})-im\zeta}.
$$
Taking the integral and introducing a new numerical
constant, $C_b=c_be^{-\e_m
m^2/4\eta+i\pi/4} \sqrt{\pi/\eta}$,  we obtain a
non-axisymmetric
generalization of solution found in \cite{Ziolkowski85}
\be
    u_b=C_b \sqrt{\det \G(\b)}
        \exp
        {i m S_b},\qquad
         S_b =   \frac{\theta\eta}m -\frac{\b m}{4{\eta}} ,
    \label{KGFbeam}
\ee
and $\theta$ is defined in \Ref{Theta}.

The absolute value of this solution does not depend on
$\a$
\be
   \left| u_b \right| = \left|C_b \sqrt{\det \G(\b)}
    \right| \exp{-\eta (\br_\perp,\Im\G(\b) \br_\perp)}.
    \label{absbeam}
\ee
The level surfaces of (\ref{absbeam}) are moving with a
unit
velocity along the $z$--axis in the negative
direction, and $\left| u_b \right|$ is exponentially
localized in transversal directions provided the matrix
$\G$ satisfies the  conditions discussed in the
Section \ref{Wave-Equation}.

As it follows from    (\ref{GammaSol}), $\G(\b) \sim
\b^{-1}{\bf E} - \b^{-2}\G_0^{-1}$ for $|\b|\to\infty$,
and thus
localization degree around $z$--axis is decreasing with
time and the solution becomes more axisymmetric. Apart
from this, the absolute value of the pre-exponential
factor is also
decreasing as $|\b|^{-d/2}$   with $|\b|\to\infty$.
Thus, at every given moment of time the solution
(\ref{KGFbeam}) has
a Gaussian localization along the transversal coordinates
and
power-law localization along the longitudinal ones.
The total energy
of the beam is infinite. Thus, the considered solution
is indeed a Gaussian
beam, one can also call it a Focus Wave Mode for the Klein--Gordon-Fock equation.



\subsection{Gaussian packets for the
Klein-Gordon-Fock equation \label{loc-part-KGF}}

Acting along the lines of Subsection
\ref{Wave-Equation}.\ref{Part_WE} we seek the particle--like
solutions of the KGFE in the form of a superposition
of the beam--like solutions $u_b$
\be
    u^{(\nu)}_p (\a,\b,\br_\perp) = \int_0^\infty d\eta\,
    F^{(\nu)}(\eta) u_b(\a,\b,\br_\perp; \eta).
    \label{superposKGF}
\ee
 By choosing $F^{(\nu)}(\eta)$ as
(\ref{Fnu}) and plugging it into (\ref{superposKGF}),
we immediately
recognize the same integral representation for the
modified Bessel
function (\ref{tabl-in}) with
$a = (4 \g \k^2 + \e_m m^2 + i \b m^2)/4$ and
$b = \g - i \theta$, and arrive at
\be
    u_p^{(\mu-1/2)}={C_p} \sqrt{\det \G(\b)}
        \( \frac{S_p}{\tau + i \b } \)^{\mu}
        K_{\mu}\( m S_p \),
        \label{KGFpacketS}
\ee
where we use notation for the complex phase function
\be
    S_p =  \[(\g - i\theta)(\tau + i\b)\]^{1/2},
    \label{S_p}
\ee
here $\tau=4 \g \k^2/m^2+\e_m$,
$C_p= 2^{\mu+1} \sqrt{\pi} e^{i\pi/4} c_b/m^{\mu} $ and
$\mu=\nu+1/2$ can be treated
as new independent parameters. %
The square root with the positive real part is assumed in
(\ref{S_p}).
The constructed family of solutions $u_p$ is
nonaxisymmetric mutlidimensional generalizations
of the solutions obtained for $d=2$ and $d=3$ in
\cite{PeFi01} .

We stress here that \Ref{tabl-in} is indeed applicable
for \Ref{superposKGF}.
First, $\Re{a}$ is positive since $\tau>0$ and $\b$
is real.
Secondly, $\Re{b} = \g + \Im{\theta}$ is positive as well,
because $\g>0$ and $\Im{\theta}>0$, as it follows from
its definition (\ref{Theta}) and the fact that the imaginary
part of the quadratic form
$({\br_\perp},\G(\beta) {\br_\perp })$ is positively
defined by assumption.


We note that for $m \to 0$ this solution transforms to
the localized
solution of the wave equation
$\phi_p^{(\mu)}$  (\ref{wavePa}) in the account that
$\tau$
changes with $m$  in such a way  that
$m^2 (\tau(m) + i \b)
\mathop\to_{m \to 0} 4 \g \k^2$.


Let us consider now some particular examples of the
constructed  solutions. The modified Bessel function
reduces to elementary functions if $\mu$ is half-integer.
For the values $\mu = 1/2$ and $\mu = -1/2$, we have
$
K_{\pm 1/2}\(  m S_p \) = \sqrt{{\pi}/{(2m S_p)}}
\exp{- m S_p}
$
and the formula (\ref{KGFpacketS}) yields
\be\label{simKGF2}
    u^{(0)}_p = C_2 \sqrt{\det \G(\b)}
         \frac{e^{-mS_p}}{(\tau + i\beta)^{1/2}}, \qquad
    u^{(-1)}_p = C_3 \sqrt{\det \G(\b)}
        \frac{e^{-mS_p}}{(\g - i\theta)^{1/2}},
\ee
 where $C_2$ and $C_3$ are numerical  constants.
We further note, that in two-dimensional space
(i.e. with  $\G\equiv \Gamma_{xx}= (\b-i\tau)^{-1}$)
we can reduce the second solution of (\ref{simKGF2})
to a function depending on one variable $S_p$ only
\be
    u^{(-1)}_p (S_p) = C_3 \frac{e^{-mS_p}}{ S_p} .
    \label{simpleKGF}
\ee
In this case, the phase $S_p$   (\ref{S_p}) is given by
\be
S_p = \sqrt{ (\g-i\a)(\tau+i\b) + x^2} =
\sqrt{x^2+(z - i e)^2  - (t - if)^2},
\ee
with $e=(\tau-\g)/2$, $f=(\tau+\g)/2$. We note that such
$S_p$ can be interpreted as a distance in the (euclidian)
space--time with imaginary time
$$
    S_p= |{\bf R}|, \qquad
    {\bf R}=(x, z-z_0, it-t_0),
$$
where $z_0=ie$, $t_0=- f$.  From this point of view,
(\ref{simpleKGF}) can be thought of  as a point source
solution $G({\bf R})\equiv u^{(-1)}_p(|{\bf R}|)$ of
the equation
\be\label{Helm}
    \Delta_3 G - m^2 G = \delta({\bf R}),
\ee
where $\delta({\bf R})$ is the three--dimensional  Dirac
delta--function, $\Delta_3  $ is the Laplacian in three
dimensions.  So, our solution of the KGFE in two spatial
dimensions is the point source solution of the elliptic
equation (\ref{Helm}) in three dimensions which is
analytically continued  to the complex plane.


Finally, the solutions \Ref{KGFbeam} and \Ref{KGFpacketS}
may be treated from the point of view of ray method approach
discussed for KGFE by Maslov \cite{Maslov}. Phase
functions  $S_b$ and $S_p$ satisfy the eikonal equation
\be\label{eikon-KGFE}
    (\partial_t S)^2 - (\nabla S)^2 - 1 =0
\ee
and solutions $u_b$ and $u_p$ may be regarded as ray
expansions
$
    u = e^{imS(\br,t)} \,\sum_{k\ge0}  {(im)^{-k}}{g_k(\br,t)}
$
 comprising  a single term.
\section{ Asymptotic  investigation
in space--time domain}\label{AsyInv_sect}

We prove here that formula (\ref{KGFpacketS}) gives a
family of particle-like solutions
of the Klein-Gordon-Fock equation.
To do so, we first note that in all cases when
$|S_p|\to\infty$, $-3\pi/2<\arg S_p< \pi/2$ we can use
the following  asymptotic expression of the modified
Bessel function
\be\label{as-Bes}
    K_\mu(m S_p) \mathop{\simeq}_{|S_p|\to\infty}
    e^{-m S_p}\sqrt{\frac{\pi}{2m S_p}} \(1+O(1/S_p)\).
\ee
Then all the solutions of our family behave as
\be
u_p^{(\mu-1/2)} \mathop{\simeq}_{|S_p|\to\infty}
     C_p \frac{\pi^{1/2}}{(2m)^{1/2} }  \sqrt{\det \G(\b)}
    \frac{S_p^{\mu - 1/2} }{(\tau + i \b )^{\mu}}
         \exp{- mS_p }.
    \label{pack-estim2}
\ee
Basing on this expression we will develop in what follows
asymptotical expansions of  \Ref{KGFpacketS}.

\subsection{Behaviour at spatial infinity}\label{moderT}
We show now that the solution  (\ref{KGFpacketS})
has exponential
decay at spatial  infinity, i.e. for $z$,
$\br_\perp\to\infty$ and finite times, and therefore has
a finite
energy.

To prove the applicability of expression (\ref{pack-estim2})
for  $z, \br_\perp\to\infty$ and fixed  time, first we give
the estimate from below for the absolute value of $S_p$
\be
     |S_p| =|\(\tau + i \beta\)\(\g - i(\a+(\br_\perp,
     \G \br_\perp))\)|^{1/2}
        \ge (\tau^2 + \b^2)^{1/4}
        (\g + |\br_\perp|^2 h(\b) )^{1/2}=
    \label{absS2}
\ee
$$
      ((\tau^2 + \b^2) (\g^2 + 2\g|\br_\perp|^2
      h(\b)+|\br_\perp|^4 h^2(\b)  ))^{1/4}
    \ge (  \b^2 \g^2 +Q |\br_\perp|^2 )^{1/4}.
$$
At the fisrt line of (\ref{absS2}) we have used that
$|\g - i \theta| \ge \Re{(\g - i \theta)} =
\g + {(\br_\perp, \Im\G \br_\perp)}$ and introduced the
notation
$$
    h(\b) = {\(\frac{\br_\perp}{|\br_\perp|},
    \Im\G \frac{\br_\perp}{|\br_\perp|}\)},
$$
while to proceed to the second line we further used that
\be\label{Gam-b1}
    \G
     \equiv   \frac{\b^{-1}}{ {\bf E} + \b^{-1} \G_0^{-1} }
     = \b^{-1}{\bf E} - \frac{\b^{-2} \G_0^{-1}}{ {\bf E}
     + \b^{-1} \G_0^{-1} }
\ee
and the fact that the imaginary part of $\G$ is positively defined.
Together with the   continuity and boundedness
of $h$ as a function of $\b$ and $\br_\perp$, it allows
us to
conclude  that there exists a constant $Q>0$ such that
for any $\b$
and $\br_\perp$
$$
    2 \g(\tau^2 + \b^2)h(\b) \ge Q.
$$
This justifies the last inequality of  (\ref{absS2})
and thus proves that $|S_p|$ does indeed grow with
$z\to\infty$ and/or $\br_\perp\to\infty$.

On the other hand, since $\Re {(\tau + i \beta)}>0$
and $\Re{(\g - i\theta)}>0$  for any finite $\b$ we have
 $ |{\arg(\tau + i \b)}|<\pi/2-\delta,$
$ |{\arg(\g - i \theta)}|<\pi/2, $
where $\delta\equiv\delta(\b)>0$ and thus
\be\label{argS1}
    |\arg{S_p}|=
    |\arg{\((\tau + i \beta)(\g - i\theta)\)^{1/2}}|
    < \pi/2-\delta.
\ee
This already proves applicability of the asymptotical
expression  (\ref{pack-estim2}). Furthermore,  assuming
that  $|z|$ is big enough we deduce that
 $   {\rm sgn}\({\arg(\tau + i \b)}\) = {\rm sgn}{(z)},$
while
$$
 {\rm sgn}\({\arg(\g - i \theta)}\) =
 -{\rm sgn}\({(\a+\Re{(\br_\perp, \G \br_\perp)})}\)
 \simeq -{\rm sgn}\((z+|\br_\perp|^2/z)\)
    =  -{\rm sgn}z,
$$
where we again used (\ref{Gam-b1}). This tells us that
for $z\to\infty$, the arguments of the factors
$\tau + i \b$ and $\g - i \theta$ in $S_p$ have opposite
signs 
and (at least partially) cancel each other. Then the
estimate
(\ref{argS1}) can be further strengthen \be\label{argS2}
    |\arg{S_p}|=|\arg{\((\tau + i \beta)
    (\g - i\theta)\)^{1/2}}| < \pi/4.
\ee

The estimates (\ref{absS2}),   (\ref{argS2}) and the
formula (\ref{pack-estim2}) show that the absolute value
of any solution
$|u^{(\mu-1/2)}_p |$ decreases exponentially with growing
coordinates, and therefore all these solutions have finite
energy.


\subsection{Asymptotics for small and moderate $z$ and time $t$}
We intend to show now that for some relation  (to be discussed below) between the mass $m$ and the solution parameters $\g$, $\tau$
the   solution (\ref{KGFpacketS}) is a wave packet
with the Gaussian envelop moving with group speed.

First we assume that the coordinates and time are small
enough (in what follows we clarify the formal meaning
of the smallness) for the square root in (\ref{S_p}) to
be expanded in   Taylor series up to the terms of the
second order in time and coordinates
\be\label{S-mod-t1}
S_p = m \sqrt{\g \tau} \left( 1+ \frac{i \b}{2\tau}
      -\frac{\b^2}{8\tau^2}
     +\ldots \right)
    \left( 1 - \frac{i\theta}{2\g} + \frac{\alpha^2}{8\g^2}
            +\ldots \right).
\ee
We use $\alpha$ instead of $\theta$ in the last term because
we are interested only in the terms quadratic in coordinates
and time. However, we postpone expanding $\G(\b)$ till we
work out the applicability conditions.
Simplifying (\ref{S-mod-t1}) and collecting the terms we
obtain
\be\label{S-mod-t}
    S_p =  \sqrt{\g \tau} \( 1 - i \frac{(\tau-\g) z -
    (\tau+\g) t }{2\g\tau}  +
         {(z - v_{gr} t)^2}
         \frac{(\g+\tau)^2}{8(\g\tau)^{2}}   -
        \frac{i }{2\g}(\br_\perp,  \G(\b)  \br_\perp) +
        \ldots \),
\ee
\be
    v_{gr}=\frac{\tau-\g}{\tau+\g}.
    \label{v_gr}
\ee
We choose for definiteness that $\g<\tau$. It  ensures
that the solution propagates forward along the $z$--axis,
i.e. that $v_{gr}>0$.

We  notice now that the expansion (\ref{S-mod-t1})
requires the conditions
$$
{|\b|} \ll {\tau} ,\qquad
    {|\theta|} <     {{|\a|+|({\bf r}_\perp,\G(\b)
    {\bf r}_{\perp})|}  }\ll  {\g}
$$
which are reduced under assumption that $\g<\tau$
to the following (actually, stronger) ones
\be\label{cond-t-mod-3}
  {|t|} \ll {\tau},\quad
     {|z-t|} \ll {\g},\quad
     {|({\bf r}_\perp,\G(\b) {\bf r}_{\perp})|}
     \leq {\bf r}_\perp^2 \| \G(\b) \| \ll {\g}
\ee
where $\|\cdot\|$ is an appropriate matrix norm,
e.g. euclidian.

The  possible expansion of the $\G(\b)$ depends on
the range of values of $t$.
First we note that if
\be
    \left\|\G_0\right\|  \ll \tau^{-1}
    \label{G0_cond}
\ee
then
 $   | z+t |\left\|\G_0\right\|
          \ll 1,$
and we can expand $\G(\b)$ in the following way
\be\label{hel-Lam}
\G(\b)\equiv{\G_0}({\bf E}+\b \G_0)^{-1} =
    {\G_0}\(1-(z+t)\G_0+\ldots\).
\ee
The same is true if (\ref{G0_cond}) is not satisfied but
$t$ is small enough for the condition $   | z+t |\left\|\G_0\right\|
          \ll 1,$
  to be fulfilled.
If both conditions are not meet, we can still expand
$({\bf r}_\perp,\G(\b) {\bf r}_{\perp})$ by using
\be\label{hel-Lam}
\G(\b) =
    \frac{\G_0}{({\bf E}+2 t \G_0)+\a \G_0}
    = \frac{\G_0}{({\bf E}+2 t \G_0)}
    \(1-\frac{\a\G_0}{{\bf E}+2 t \G_0}+\ldots\)
\ee
the latter expansion is valid if
\be
    \left\|{(z-t)\G_0}\({\bf E}+2 t \G_0\)^{-1}\right\|
    \ll \g \|\G(2t)\| \ll 1.
    \label{cond-t-mod-1}
\ee
Thus, the dependence on $t$ can be important even for small
times, $t \ll \tau$, if the  $\|\G_0\|$ is large enough.

Thus we conclude, that if the conditions
\Ref{cond-t-mod-3} and
either one of the \Ref{G0_cond} and
\Ref{cond-t-mod-1} are valid,
then we may use the expansion \Ref{S-mod-t} of $S_p$ and
the asymptotics  (\ref{as-Bes})  of the modified Bessel
function to
obtain the asymptotics of the packet (\ref{KGFpacketS})
\be\label{pack-mod-t}
u_p^{(\mu-1/2)} \mathop{\simeq}_{p\to\infty}
            A(\zeta)  \exp{ i ({\cal K} \tilde{z} -
            \Omega \tilde{t} +  g |\tilde{\br}|^2_{\perp})}
        \exp{ - \frac{(\tilde{z} - v_{gr}
        \tilde{t})^2}{2\Delta_{\parallel}^2}
            -  \frac{|\tilde{\br}_\perp|^2}
            {2\Delta_{\perp}^2}},
\ee
where we used the dimensionless coordinates,
time and mass
\be\label{non-dim}
\tilde{t} = {t}/{\sqrt{\tau \g}}, \quad \tilde{z} =
{z}/{\sqrt{\tau \g}}, \quad \tilde{\br} =
{\br}/{\sqrt{\tau \g}}, \quad p = m \sqrt{\tau \g}.
\ee
All the characteristics of the asymptotic ---
${\cal K},$ $\Omega,$ $v_{gr}$, $\Delta_{\parallel}$
and $\Delta_{\perp}$ in (\ref{pack-mod-t}) are expressed
in terms of  non-dimensional parameters $p$,
$\tau/\g$ and $\tau \Im{\G}$
as follows
\be
\O = {p} \( \sqrt{{\tau}/{\g}} + \sqrt{{\g}/{\tau}} \)/2,
\quad
    \K = {p} \( \sqrt{{\tau}/{\g}} -
    \sqrt{{\g}/{\tau}} \)/2, \quad v_{gr} = {\K}/{\O},
    \label{def-o-k}
\ee
\be \label{pack-mod-t-ds}
   \Delta_{\parallel}^2  
    = \frac{p}{\Omega^2}, \quad\qquad
     \Delta_{\perp}^2 =
     \frac{1}{p \tau({\bf e}_\perp,\Im\G(\zeta)
     {\bf e}_{\perp})} , \quad
    \zeta=\begin{cases} 0,&|t |\left\|\G_0\right\|
    \ll 1\\ 2 t, &\rm otherwise \end{cases}
\ee
here ${\bf e}_\perp={\br_\perp}/{|\br_\perp|}$.
The amplitude factor $A$ and correction term in
the phase $g$ read
\be\label{pack-mod-t-A}
A=C_p \frac{\pi^{1/2}}{(2m)^{1/2}  }
\sqrt{\det \G(\zeta)} \,
    \frac{\g ^{\mu/2 - 1/4} }{\tau^{\mu/2+1/4}}
    \exp{-m \sqrt{\g \tau}}, \quad
  g = \frac{p}{2}\tau({\bf e}_\perp,\Re\G(\zeta)
  {\bf e}_{\perp}).
\ee
The solution $u_b$ describes a wave with frequency
$\Omega$ and wave number ${\cal K}$, which propagates
along the $z$
axis  and has the Gaussian envelop moving with group
velocity $v_{gr}$. The  localization near the $z$ axis
is determined by the ${\Delta}_{\perp}$ which depends
on the orientation of $\br_\perp$ (the result of
astigmatic 
nature of the considered solution) and time.
 Note that for $t> \|\G_0\|^{-1}$ (while still being
 much less then $\tau$) the width ${\Delta}_{\perp}$
 starts growing linearly with time. 
We call this regime as moderate times' one. It can only show up when $ \|\G_0\|\ll \tau$, otherwise only two asyptotic regimes can be identified for our solution: small times or large ones.


 Now we will check, that the formulae
 \Ref{pack-mod-t}-\Ref{pack-mod-t-A} describe correctly
 the field up to the distances where the packet becomes
 exponentially small.
  To this end we first compare (by order of magnitude)
  the longitudinal width of the packet,
  $   \Delta_{\parallel}\sim |\tilde{z}-v_{gr}\tilde{t}| $,
with the distance from the point
$\tilde z_0=v_{gr}\tilde{t}$, where (\ref{pack-mod-t})
is still applicable as defined by the second condition
(\ref{cond-t-mod-3}). This distance is
$|\tilde{z}-v_{gr}\tilde{t}|\le |\tilde{z} -
\tilde{t}| + |\tilde{t}(1-v_{gr})| \ll \sqrt{\g/\tau}$
with account of  \Ref{v_gr}.
Secondly,  the transverse width of the packet
$|\tilde{\br}_\perp| \sim \Delta_{\perp}$ should be inside
the zone determined by the third condition of
(\ref{cond-t-mod-3}).  Thus, it must hold that
\be\label{cond1-2}
    \Delta_{\parallel} \ll \sqrt{{\g}/{\tau}},
    \qquad \Delta_{\perp} \ll (\tau \| \G(\zeta) \|)^{-1/2}.
\ee
Thirdly, we demand that our solution must travel according
to (\ref{pack-mod-t}) on distances which are much
larger then its longitudinal width $\Delta_{\parallel}$,
i.e.,
\be\label{cond-3}
    \Delta_{\parallel} \ll \tilde{z}_{max}\sim v_{gr}
    \sqrt{{\tau}/{\g}}.
\ee
The two widths $\Delta_{\parallel}$ and $\Delta_{\perp}$
contain $p$ in the denominator. Therefore all of the
conditions (\ref{cond1-2}) and (\ref{cond-3}) are satisfied
if $\tau/\g$ is fixed and $p \to \infty$. The last
condition (\ref{cond-3}) is the more
restrictive. In terms of the parameters $\g/\tau$ and $p$
it reads
 \be\label{mod-t-cond-t}
   \frac{1}{p} \ll \frac{\tau}{4\g}\(1 -
   \frac{\g}{\tau}\)^2,{\rm  i.e.\ } p \ll {\cal K}^2 \frac{\Omega+{\cal K}}{\Omega-{\cal K}}.
\ee
Finally, we may compare our  results in the limit
$m \to 0$ with the formulas for the packet--like
(stigmatic) solution for the wave equation
obtained in (37) of \cite{PerSid}.  In doing so we must
put  $m^2 \tau \to 4 \g \kappa^2$, where $\kappa$ is a
constant used in \cite{PerSid} (compare with the note
after (\ref{KGFpacketS}))  and also
$\G_0 = i \varepsilon^{-1}{\bf E}$ and thus obtain
$$
 \Delta_{\parallel} = \frac{4(\g\tau)^{3/2}}{m(\g+\tau)^2}
 \to \frac{2 \g}{\kappa}, \quad \Delta_{\perp} =
 i \frac{\sqrt{\g}}{m\sqrt{\tau}}(-i\varepsilon) \to
 \frac{\varepsilon}{2\kappa}
$$
in complete agreement with \cite{PerSid}.

\subsection{Large-time behavior}\label{largeTsec}

Let us find the asymptotics of   (\ref{KGFpacketS}) for
large times and large distances ${\br}={(\br_\perp,z)}$.
We assume that  ${\br} = {\bv} t$ for  $t \to \infty$,
but ${\bv} =(\bv_\perp,v_z)$ is fixed. The  asymptotics
of $\G$   (\ref{GammaSol}) and $\theta$   (\ref{Theta})
are as follows
\be\label{help-t}
    \G 
    \mathop{\simeq}_{t\to\infty}\frac{\bf  E}{(v_z+1)t}
    -\frac{\G_0^{-1}}{(v_z+1)^2t^2} +O(t^{-3}),
    \qquad
\theta  \mathop{\simeq}_{t\to\infty} -t
\frac{1-v^2}{1+v_z} -
        \frac{(\bv_\perp, \G_0^{-1}\bv_{\perp})}
        {(v_z + 1)^2} + O(t^{-1}),
\ee
here we used that $\b=(1+v_z) t$. Then the  $S_p$
(\ref{S_p}) can be expanded as
\be\label{as-s-help}
    S_{p}=   (\theta \b)^{1/2}\left(1 + i
    \frac{\g}{\theta}\right)^{1/2}
        \(1 - i\frac{\tau}{\b}\)^{1/2}
\mathop{\simeq}_{t\to\infty}  (\theta \b)^{1/2}
\(1 +  i \frac{\g}{2\theta} - i \frac{\tau}{2\b}
+ O(t^{-2})\)
\ee
\be
= i t\sqrt{1 - v^2} +  \frac{i (\bv_{\perp},\G_0^{-1}
\bv_{\perp})}
    { 2 (1+v_z)\sqrt{1 -v^2} } +
     \frac{     \g (1+v_z)  }{ 2  \sqrt{1 - v^2} }
       + \frac{\tau\sqrt{1 - v^2}}{2(1+v_z)}
+O(t^{-1}) .
             \label{S_p_as}
\ee
In the first line here we took the square root with the
positive
real  part. Thus, $|S_p| \to \infty$ for $t \to \infty$
and we can
use the asymptotics (\ref{as-Bes}) of $K_{\mu}(mS_p)$,
and  (\ref{pack-estim2}) for the whole 
solution.

Now,  we suppose that $p \equiv m\sqrt{\g\tau}\gg 1$.
Introducing for  the sake of brevity new variables
\be\label{no-nk}
    \no=1/\sqrt{1 - v^2}, \quad
    {\bm \chi}={\bv}/\sqrt{1 - v^2},
\ee
we rewrite the solution (\ref{pack-estim2}) in the
form that allows for its further
simplification
\begin{eqnarray}
&& u_p^{(\mu-1/2)} \approx {\cal A}(t, \bv)
\exp{-imt\sqrt{1 - v^2}}
    \exp{-p\Phi(\bv)},   \label{t-as-pack}\\
&& \Phi 
    = i\frac{|\nk_{\perp}|^2}{\sqrt{\g\tau}}
    \frac{({\bf e}_{\perp},\G_0^{-1}
    {\bf e}_{\perp})}{ 2 (\no+\nk_z) } +
     \frac{    \sqrt{\g/\tau} (\no+\nk_z)  }{ 2   }
     + \frac{\sqrt{\tau/\g} }{2(\no+\nk_z)},
     \quad {\bf e}_{\perp}=
     \frac{{\bm \chi}_{\perp}}{|{\bm \chi}_{\perp}|},
     \label{Phi}\\
&&  {\cal A}(t, \bv)
    \mathop{\simeq}_{t\to\infty} C_p
    \sqrt{\frac{\pi}{2 m}}\,\, e^{i\delta_{\G}-i(1+d)
    {\rm sgn}(t)\pi/4}
            \frac{(1 - v^2)^{\mu/2-1/4}}{ |t|^{(d+1)/2}
            (1+v_z)^{\mu+d/2}},\label{A_large_t}
\end{eqnarray}
which we obtained by using  that $S_p^{\mu-1/2}\approx
e^{i(\mu-1/2){\rm sgn} (t) \pi/2}
\(|t|\sqrt{1-v^2}\)^{\mu-1/2}$ for large $|t|$ due to
$S_p \approx +0 +it\sqrt{1-v^2}$.
We used also that $\Re{(\tau+i\b)}>0$, and the fact that
$\sqrt{\det(-i \G(\b))}/(\tau +
i\b)^{\mu}\simeq (i\b)^{-d/2 - \mu }
    = |\b|^{ - \mu - d/2} e^{-i(\mu+d/2)
    {\rm sgn}(t)\,\pi/2 }$.
To get the latter equality we note that for  $\b \to \infty$
we get $(-i\G)\approx {\bf E}/\({i\b+O(1)}\)$,
where $O(1)$ is positive. The branch of the square root  $\sqrt{\det{(-i\G(\b))}}$ is fixed by the asymptotics for $|t|\to \infty$: $\arg{\sqrt{\det{(-i\G)}}} \to -{\rm sgn} (t)d\pi/4$.
 We do not specify the branch of the square
root in $\sqrt{\det \G}$ and introduce the argument
 \be\label{delta-G}
 \delta_{\G}= \arg{\sqrt{\det{\G}}} -
 \arg{\sqrt{\det{(-i\G)}}}.
 \ee

When $p\gg1$, the modulus of the second exponent
in (\ref{t-as-pack}) has a sharp maximum and we will use  quadratic
approximation of $\Phi$ in its vicinity.
 We  seek  its position in
 the spherical coordinate system, i.e.,
 $\nk_z=\nk \cos{\vartheta}$,
$ {\bm \chi}_{\perp} = \nk\,$ $ {\bf e_{\perp}}
\sin{\vartheta}$.
We note that the first term in $\Phi$ \Ref{Phi}
has nonnegative real part,
 thus its least value is equal to zero when
 ${\vartheta} = 0$  or
 ${\vartheta} = \pi$.
The two other terms of $\Phi$ are mutually inverse (up to a factor of $1/2$),
thus their sum reaches its least value equal to unity if
\be
    \no+\nk_z=\sqrt{\tau/\g}.
    \label{no}
\ee
Together with the condition $\vartheta=0$ this gives us
\be\label{extrem}
{\bf v}_\perp=0,\quad v_z=v_{gr},\qquad \no =
\Omega/m, \quad \nk
        = {\cal K}/m,
\ee
where ${\cal K}$ and $\Omega$ are defined in
\Ref{def-o-k}.
It is easy to check that  $\vartheta=\pi$  is
incompatible with \Ref{no} for $v<1$, $\tau/\g>1$.
Finally,
we obtain
\be\label{lar-t-pack}
u_p^{(\mu-1/2)} \approx
    {\cal A}(t, \bv_{gr}) 
    \exp{-m\sqrt{\g\tau}-imt\sqrt{1 - v^2} -
    i p \vartheta^2\frac{\Im{\Phi''_{\vartheta\vartheta}}}
    {2}}
        \exp{- \frac{(\br-\bv_{gr}t)^2}{2t^2\Delta_v^2} -
        \frac{\vartheta^2}{2\Delta_{\vartheta}^2} }.
\ee
We have used here that $v_z=z/t$,
$\bv_{\perp}=\br_{\perp}/t$ and
that the derivative $\Phi_{\vartheta \nk}''$  is zero.
The notation for
${\cal A}(t, \bv)$ was introduced in  \Ref{A_large_t}.
The widths of the
packet $\Delta_v$ and $\Delta_{\vartheta}$ are expressed
through the
second derivatives of $\Phi(\bv)$ as follows
\begin{eqnarray}
&&\Delta_v^2(t)
    =  \[p\Phi''_{\nk\nk}
            ( \nk'_v)^2\]^{-1}
    = \frac{(1-v_{gr}^2)^2}{p}
    = \frac{p^3}{\Omega^4},
  \label{lar-t-Del1}  \\
&&\Delta_{\vartheta}^2(t)
    =  \(p \Im \Phi''_{\vartheta\vartheta}  \)^{-1}
    =   \frac{\tau (1-v_{gr}^2)}{p\,v_{gr}^2
    ({\bf e}_{\perp},-\Im\G_0^{-1} {\bf e}_{\perp}) } =
    \frac{p}{{\cal K}^2}
    \frac{\tau}{({\bf e}_{\perp},-\Im\G_0^{-1}
    {\bf e}_{\perp})}, \label{lar-t-Del2}
\end{eqnarray}
where $\Phi''_{\nk\nk}\equiv\frac{\partial^2 \Phi}
{\partial
\nk^2}|_{\vartheta=0, v = v_{gr}}$,
$\Phi''_{\vartheta\vartheta}\equiv\frac{\partial^2 \Phi}
{\partial
\vartheta^2}|_{\vartheta=0, v=v_{gr}}$,
$\nk'_v=\frac{d\nk}{dv}|_{v=v_{gr}}$.  It is worth
mentioning here,
that due to the fact that $\Im \G_0$ is positively
defined, so is
$-\Im \G_0^{-1}$, and thus $\Delta_{\vartheta}^2$
is positive.

According to (\ref{lar-t-pack}) the field is concentrated
in the intersection of a cone and a spherical annulus.
The width of the annulus increases with time and
 may be estimated as $2t \Delta_v$. We will
require that the speed of the packet center exceeds the
speed of the packet enlarging.
The angle of the cone does not depend on time and we can
assume that it  is small. Finally we have conditions
\be\label{cond-loc}
    \Delta_v \ll v_{gr}, \quad  \Delta_{\vartheta} \ll 1,
\ee
which can be written in the simplest stigmatic case,
$\G={\bf E}/(\b-i\varepsilon)$, as follows
\be
p \gg {16}{\( {\tau}/{\g} - {\g}/{\tau} \)^{-2} },
\quad p \gg {4}{\( \sqrt{{\tau}/{\g}} -
\sqrt{{\g}/{\tau}}\)^{-2}} {\tau}/{\varepsilon}.
\ee
These conditions are the more restrictive  the closer are
$\tau$ and $\g$ to each other. If $p \to \infty$ for fixed
other parameters the localization is more pronounced.
If $\g/\tau \to 0$ the localization both in angle and
along the propagation direction is better.  In the case
of general astigmatism the term
$({\bf e}_{\perp},-\Im\G_0^{-1} {\bf e}_{\perp})$ should
stand in the last inequality instead of $\varepsilon$.

Now we turn to the applicability conditions of the obtained
formulas.
Time $t$ will be considered  large if expansions of $\G$
 and  $\theta$ (\ref{help-t}) could be limited to their
 first terms. For that we will require
\be
    |t|\,  \gg {\|\G_0^{-1}\|}/{|1+v_z| },\quad |t| (1-v^2)
    \gg {\bv_\perp^2 \| \G_0^{-1}\|}/{|1+v_z|}.
    \label{Cond1}
\ee
For the expansion of $S_p$
 (\ref{S_p_as}) being valid we additionally  need
\be
    \frac{\theta}{\g}\mathop{\simeq} \frac{|t| (1-v^2) }
    {\g|1+v_z|} \gg 1,\qquad
    \frac{\b}{\tau} = \frac{|t(1+v_z) |}{\tau} \gg 1.
    \label{Cond2}
\ee
Both conditions (\ref{Cond1}) and (\ref{Cond2}) contain $v$,
but we can substitute it with $v_{gr}$  by recalling that
$|v - v_{gr}| \mathop{\simeq} \Delta_v \ll v_{gr}$
according to (\ref{cond-loc}). The  group velocity
itself 
can be expressed in terms of $\tau$ and $\g$ by using
(\ref{v_gr}).
Two conditions (\ref{Cond2}) are reduced to just one then,
$|t| \gg\tau$.  Taking into account that $\bv_{\perp}^2
\mathop{\simeq}
v_{gr}^2 \vartheta^2 \ll 1$ we replace (\ref{Cond1})
by a stronger
inequality. Combining the two, we obtain
\be\label{lar-t-defin}
|t|\,  \gg 4 {\|\G_0^{-1}\|} \tau/\g, \quad |t| \gg \tau.
\ee
These conditions specify large times.

\section{Fourier analysis}\label{Fourier-analysis}
Let us introduce a Fourier transformation relevant
to the problem in hand
\be
{\cal F}[f](\o,\bk_\perp,k_z) = \int_{\mathbb{R}^{d+2}} dt\,
d^d \br_\perp \,d z\, f(t,\br_\perp,z)\,
        e^{i(\o t- \bk_\perp \br_\perp-k_z z)},
    \label{cal F}
\ee
 here $\bk_\perp$ denotes the Cartesian components of the
wave vector, $\bk_\perp=(k_x,k_y,\ldots)$, perpendicular to
$k_z$,   $d$ is the number of transversal dimensions.
The inverse transformation is 
$$
    f(t,\br_\perp,z) = \frac{1}{(2 \pi)^{d+2}} \int_{\mathbb{R}^{d+2}} d\o\, d^d \bk_\perp\, dk_z\,
{\cal F}[f](\o,\bk_\perp,k_z)\,
        e^{-i(\o t- \bk_\perp \br_\perp-k_zz)}.
$$

Performing the Fourier transformation of
multidimensional  
WE (\ref{wave-eqN}) or KGFE (\ref{KGFn}) one obtains the
following equation in terms of   generalized functions
\be
    (\o^2-\k^2(\bk_\perp)-k_z^2 )  
        {\cal F}[f](\o,\bk_\perp,k_z) =0,
    \label{eqFour}
\ee
 here $\k^2= \bk_\perp^2 $ for WE and
 $\k^2= \bk_\perp^2+m^2$ for KGFE.

Any solution of (\ref{eqFour}) must be representable as
\be
    {\cal F}[f](\o,\bk_\perp,k_z)
    =\delta(\o^2-\k^2(\bk_\perp)-k_z^2)\, 
         \hat f( \bk_\perp,k_z)
    \label{hat f}
\ee
with a suitable well--behaved function
$ \hat f( \bk_\perp,k_z)$. It is this function $ \hat f$
which we   call the Fourier image in what follows.

Constructing particular solutions to the WE or KGFE,
we are free to choose any particular subspace of the
surface
$\o^2=\k^2(\bk_\perp)+k_z^2$ in the phase space.
For instance, in \cite{DonnZiol92} it was considered
a solution of the from
\be
    {\cal F}[f](\o,\bk_\perp,k_z)
    = \Xi_{\tilde\eta}(k)\delta\(k_z
    -(\tilde\eta-\k^2/4\tilde\eta)\)
    \delta\(\o+(\tilde\eta+\k^2/4\tilde\eta)\)
    \label{Ziolk}
\ee
where $\tilde\eta$ is an arbitrary (real)
parameter  (to avoid conflict of notation we changed the
original notation of  \cite{DonnZiol92}), and
$\Xi_{\tilde\eta}$ is an arbitrary weight function.

Now we obtain the Fourier image both for the beam--like
solutions $\phi_b$, $u_b$ and particle--like ones $\phi_p$,
$u_p$. Apart from revealing the connection of our
solutions with aforementioned ones, it  will also be
used in constructing asymptotic expansions in Appendix
\ref{Asymptotic-KGF}.

\subsection{Wave Equation}
The Fourier transform \Ref{cal F} of the solution
(\ref{waveBeam}) of the WE is given by
$$
{\cal F}[\phi_b](\omega, k)=c_b \int dt\, dz\,
d^d\br_\perp\, e^{i(\o t- k_z z-\bk_\perp\br_\perp)} \,
    \sqrt{\det\G}\,e^{i\eta(\a+(\br_\perp,\G\br_\perp))}
$$
\begin{equation}\label{phi_b_k_def}
    =c_b \int dt\, dz\, 
    e^{i\o t-i k_z z} e^{i \eta \a} I(\eta,\br_\perp),
\end{equation}
\begin{equation}\label{I}
  I\equiv  e^{i\delta_{\G}} \sqrt{\det(-i\G)}
  \int d^d\br_\perp\, e^{ -i\bk_\perp\br_\perp
  +i\eta(\br_\perp,\G\br_\perp)}
    = \(\pi/\eta\)^{d/2}   e^{-\tfrac{i}{4\eta}
    (\bk_\perp,\G^{-1}\bk_\perp)}\, e^{i\delta_{\G}},
\end{equation}
 the definition of $\delta_{\G}$ see in (\ref{delta-G}).
 We recall that all the eigenvalues of the matrix $(-i \G)$
 have positive real part, and thus the last integral is
 convergent.

Substituting the last formula into (\ref{phi_b_k_def})
and taking into account that $\G^{-1}=\G_0^{-1}+\b {\bf E}$
we have
\begin{equation}\label{phi_b_k}
{\cal F}[\phi_b](\bk_\perp,k_z)= c_b e^{ i\delta_{\G}}
\(\pi/\eta\)^{d/2}
                e^{-\tfrac{i}{4\eta}
                (\bk_\perp,\G^{-1}_0\bk_\perp)}
    \int dt\,   e^{i\o t-it(\eta+\tfrac{1}{4\eta}
    \bk_\perp^2)}
    \int dz\, e^{-iz(k_z-\eta+\tfrac{1}{4\eta}
    \bk_\perp^2)}
\end{equation}
$$
    = c_b  e^{ i\delta_{\G}}
                e^{-\tfrac{i}{4\eta}
                \(\bk_\perp,\G^{-1}_0\bk_\perp\)}
        \frac{16 \pi^{2+d/2} \eta^{2-d/2}}{\bk_\perp^2
        + 4 \eta^2}\,
                \delta\({\eta-\frac{k_z
                +\sqrt{k_z^2+\bk_\perp^2}}{2}}\)
            \delta\(\o-\(\eta+\tfrac{1}{4\eta}
            \bk_\perp^2\)\).
$$
Using the properties of  the delta functions we can rewrite
it finally as
\be {\cal F}[\phi_b](\bk_\perp,k_z) = \hat{c}_b
        \frac{e^{-\tfrac{i(\bk_\perp,\G^{-1}_0\bk_\perp)}
        {2(k_z+\o)}}}
                {\o(k_z+\o)^{d/2-1}}\,\,
        \delta\({\eta-\frac{k_z+\o}{2}}\)
        \delta\(\o-\sqrt{\bk_\perp^2+k_z^2}\),
        \label{F_b_k_fn}
\ee
$
    \hat{c}_b \equiv  c_b \pi  \({2\pi}\)^{1+d/2}
    e^{ i\delta_{\G}}.
$
We remind that $\eta>0$ is a  free parameter of our
solution, along with $\G_0$. The Fourier image is
defined now as
(assuming $\o=\sqrt{\bk^2_\perp+k_z^2}$)
\be
\hat \phi_b(\bk_\perp,k_z) = \hat{c}_b
        \frac{e^{-\tfrac{i(\bk_\perp,\G^{-1}_0\bk_\perp)}
        {2(k_z+\o)}}}
                {\o(k_z+\o)^{d/2-1}}\,\,
        \delta\({\eta-\frac{k_z+\o}{2}}\).
        \label{phi_b_k_fn}
\ee
If compared with the considerations of \cite{DonnZiol92}
(see
eq. (\ref{Ziolk})) we can see that in our case
$\tilde\eta= \eta $
and
$$
    \Xi\equiv\Xi(\bk_\perp,k_z)= \hat{c}_b
    \frac{e^{-\tfrac{i(\bk_\perp,\G^{-1}_0\bk_\perp)}
    {2(k_z-\o)}}}
                { (k_z-\o)^{d/2 }}.
$$
This shows that our spectral function  has more variables
and less symmetries  depending on $\bk_\perp$ and $k_z$
separately.

For obtaining the Fourier image of a particle like
solution $\phi_p(k)$ we employ \Ref{FnuWave}
$$
    \hat\phi_p^{(\nu)}(\bk_\perp,k_z)=
    \int_0^\infty d\eta\,
    \hat \phi_b(\bk_\perp,k_z)F^{(\nu)}(\eta)
    $$
    $$
    =
        \frac{\hat{c}_b e^{-\tfrac{ i(\bk_\perp,
        \G^{-1}_0\bk_\perp)}{2(k_z+\o)}}}
                {\o(k_z+\o)^{d/2-1}}
                \int_0^\infty d\eta\,  \eta^{-\nu-1}
                e^{-\g(\eta+\k^2/\eta)}
                \delta\({\eta-\frac{k_z+\o}{2}}\).
$$
Then,  we can write
\begin{equation}\label{phi_p(k)}
\hat \phi_p^{(\nu)}(\bk_\perp,k_z)=2^{\nu+1}\hat{c}_b
\,\frac{e^{ -\tfrac\g2 (k_z+\o)
    -\tfrac{4 \g \k^2+i(\bk_\perp,\G^{-1}_0\bk_\perp)}
    {2(k_z+\o)}}}
{\o \(k_z+\o\)^{\nu+d   /2}} .
\end{equation}


\subsection{KGF equation}

Now we shall construct the Fourier transform of the KGF
solution (\ref{KGFbeam}). We accomplish it by acting
similar to the Section \ref{KGFeqn}\ref{beamsKGF}.
First we increase the number of dimensions by one,
$d\to d+1$, and put the additional momenta component
equal to mass, $k_{d+1}\equiv m$.

Thus, we have to  substitute everywhere in
(\ref{phi_b_k_fn}) $d$ by $d+1$ and $\bk_\perp$ by
$\bk_\perp^m=(\bk_\perp,m)$
$$
    \hat u_b (\bk_\perp,k_z)\equiv \hat\phi_b
    (\bk_\perp^m,k_z)
        = \hat\phi_b ((\bk_\perp,m),k_z).
$$
At the same time we must assume similar to
(\ref{zeta}) that
$$
    (\bk_\perp^m,\tilde\G^{-1}_0\bk_\perp^m)=
    (\bk_\perp,\G^{-1}_0\bk_\perp)-i\e_m m^2,
$$
where $\tilde\G_0$ is a $(d+1)\times (d+1)$ matrix,
and $ \G_0$ is $d\times d$ one as before.

Thus,  the Fourier image of the beam--like solution is
\be
\hat u_b(\bk_\perp^m,k_z)    =    \hat{C}_b
        \frac{e^{-\tfrac{\e_m m^2+i(\bk_\perp,
        \G^{-1}_0\bk_\perp)}{2(k_z+\o)}}}
                {\o(k_z+\o)^{d/2-1/2}}\,\,
        \delta\({\eta-\frac{k_z+\o}{2}}\)
        \label{u_b_k_fn},
\ee
$
    \hat{C}_b = c_b \pi  \({2\pi}\)^{3/2+d/2}
    e^{ i\delta_{\G}}.
$
Here  it is assumed that $\o=\sqrt{m^2+\bk_\perp^2+k_z^2}$.

For obtaining the Fourier image of the particle--like
solution $u_p(\bk_\perp,k_z)$ we perform the integral
transformation  (\ref{superposKGF}) of the \Ref{u_b_k_fn}
$$
    \hat u_p^{(\nu)}(\bk_\perp,k_z)=\int_0^\infty d\eta\,
    \hat u_b((\bk_\perp,m),k_z)F^{(\nu)}(\eta)
    $$
    $$
    = \hat C_b \,\,
        \frac{e^{-\tfrac{\e_m m^2+i(\bk_\perp,
        \G^{-1}_0\bk_\perp)}{2(k_z+\o)}}}
                {\o(k_z+\o)^{d/2-1/2}}
                     \int_0^\infty d\eta\,
                     \eta^{-\nu-1} e^{-\g(\eta+\k^2/\eta)}
                     \delta\({\eta-\frac{k_z+\o}{2}}\)\,.
$$
Then we arrive at
\begin{equation}\label{u_p(k)}
\hat u_p^{(\nu)}(\bk_\perp,k_z)=\hat C_p
\,\,\frac{e^{ -\tfrac\g2 (k_z+\o)
    -\tfrac{\tau m^2+i(\bk_\perp,\G^{-1}_0\bk_\perp)}
    {2(k_z+\o)}}}
{\o \(k_z+\o\)^{\nu+d   /2+1/2}}
\end{equation}
where we used the notation of the previous section,
$\tau=4 \g \k^2/m^2+\e_m$,
and put $ \hat C_p=2^{\nu+1}\hat C_b$.

We notice a remarkable difference in localization properties of the Fourier images for KGFE and WE:
in $\hat
u_b(\bk_\perp,k_z)$ as compared with
$\hat\phi_b(\bk_\perp,k_z)$ it is
the absence the exponential suppression of small $(k_z+\om)$ via
terms of the type of $e^{-c/(k_z+\om)}$.



\section{Discussion of the results}\label{Discu_sec}

In the present paper we have elaborated  four families
of explicit exact exponentially localized solutions to
the wave equation  (\ref{waveBeam}), (\ref{wavePa}),
and to the Klein--Gordon--Fock one  (\ref{KGFbeam}),
(\ref{KGFpacketS}). The families (\ref{waveBeam}),
(\ref{KGFbeam})  represent beam--like solutions localized
exponentially near a ray, while  (\ref{wavePa}) and
(\ref{KGFpacketS})  are particle--like ones localized
exponentially near a point moving with group velocity
along one of the axis. All  of the presented solutions
are  astigmatic multi-dimensional generalizations of those
obtained before by the authors
\cite{KisPlach12,PeFi01, KiselevPerel00}, as well as by
other researches \cite{Besieres90,Besieres00}, etc.

Unlike most of the others works, we performed all the
analysis in space--time domain, which proved to be both
convenient and efficient. Focusing on the particle--like
solutions of the KGF equation which are somewhat less
studied in the literature, we investigated in detail the
asymptotic properties of the central result of our work ---
the particle--like solutions to the KGF equation
distinguishing several regimes: small times, moderate
times and large times. We also presented explicit Fourier
transformation  of all constructed solutions and confirmed
our asymptotic consideration obtained in space--time domain
by investigating the Fourier integral.

Now we summarize briefly the contents of the Section
\ref{AsyInv_sect}.
The constructed solutions contain several parameters:
$\tau,$ $\g$, $\G_0$ (or $\e$ in the stigmatic case). 
If the non-dimensional mass
\be \label{par-p}
p = m \sqrt{\g \tau}
\ee
is large the solution behaves as a packet with the Gaussian
envelop filled with oscillations which on the axis of the
packet has  the wave number ${\cal K}$ and the frequency
$\Omega= \sqrt{{\cal K}^2 + p^2}$, and moves with the
group speed $v_{gr}={\cal K}/\Omega$. Packet--like
behaviour takes place for all times.
Below we discuss properties of the field in different
regimes.

\begin{figure}
\centering
\includegraphics[width=5in]{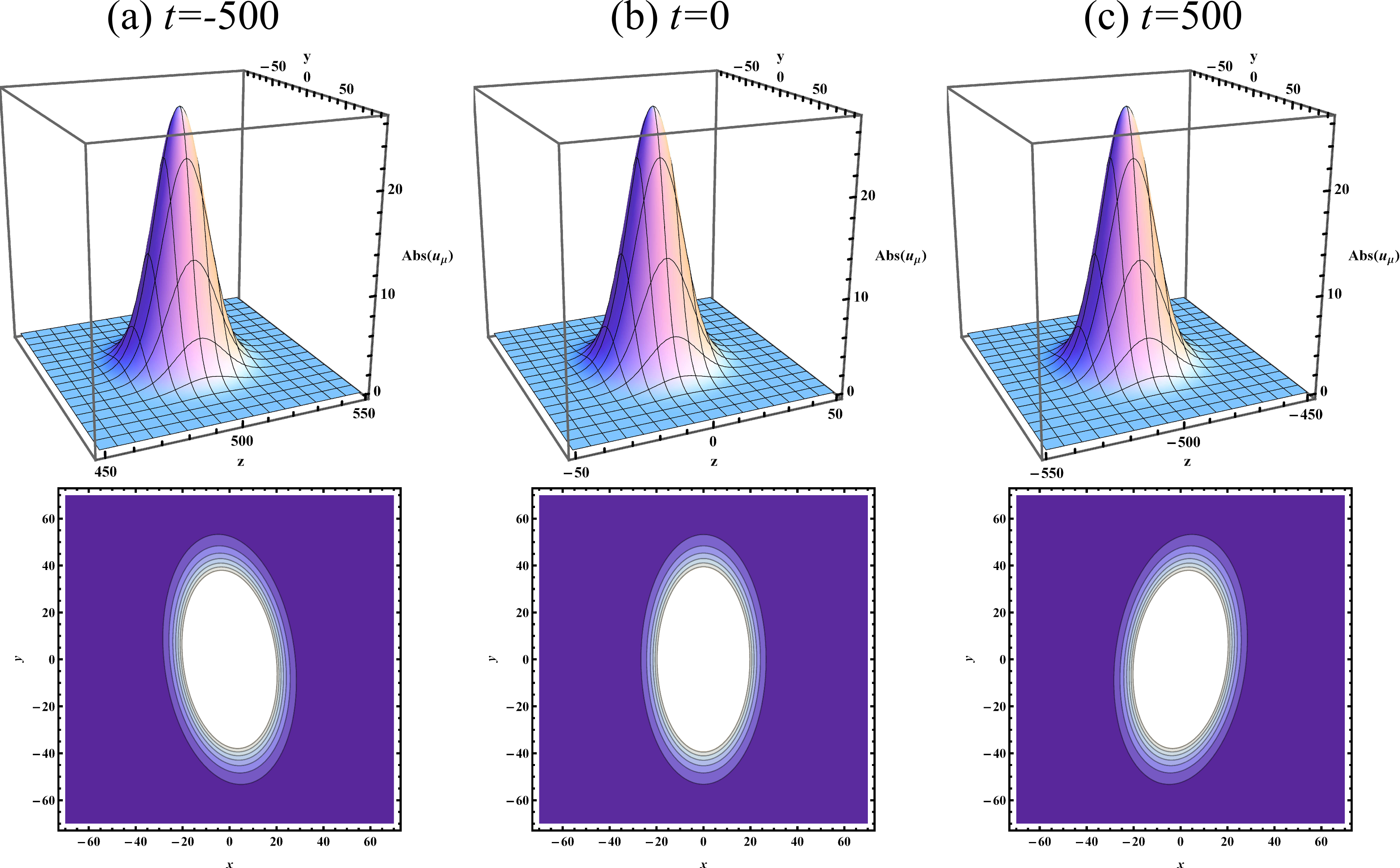}
\caption{Small  times bahviour of absolute value of $u_p$ for consequent times $t=-500,0,500$ as a function of $y$ and $z$ for $x=0$ (upper pictures) and as function of $x$ and $y$ for $z=v_{gr}t$ (lower pictures). Convenient normalization for the $|u_p|$ is chosen,  the coordinartes and time are in the units of mass $m$. See text for the values of all parameters.}
\label{small_t_pic}
\end{figure}

For small times, $t\ll \|\G_0\|^{-1}$ (assuming that
$\|\G_0\|^{-1}<\tau $) the solution behaves according
to \Ref{pack-mod-t} with $\zeta=0$. This regime is
characterized by complete absence of any distortion
during the propagation. The longitudinal width as well
as transversal one, is time independent. In the transversal
direction the astigmatic properties are practically frozen and do not
depend neither on time, nor on propagated distance. The
localization ellipse is defined by $\G_0$ itself (compare
with large times).
The maximum of the propagated distance is of the order
of $ 2v_{gr}\|\G_0\|^{-1}$. This regime is exemplified at the Fig.\ref{small_t_pic} for the following values of parameters ( in uits of mass $m$) $\g=800$, $\tau=8\cdot10^5$, $q_1=1+i \e$, $q_2=14+ 3 i \e$, $\e=3\cdot 10^3$, $\Phi=-0.31 i$.

\begin{figure}
\centering
\includegraphics[width=5in]{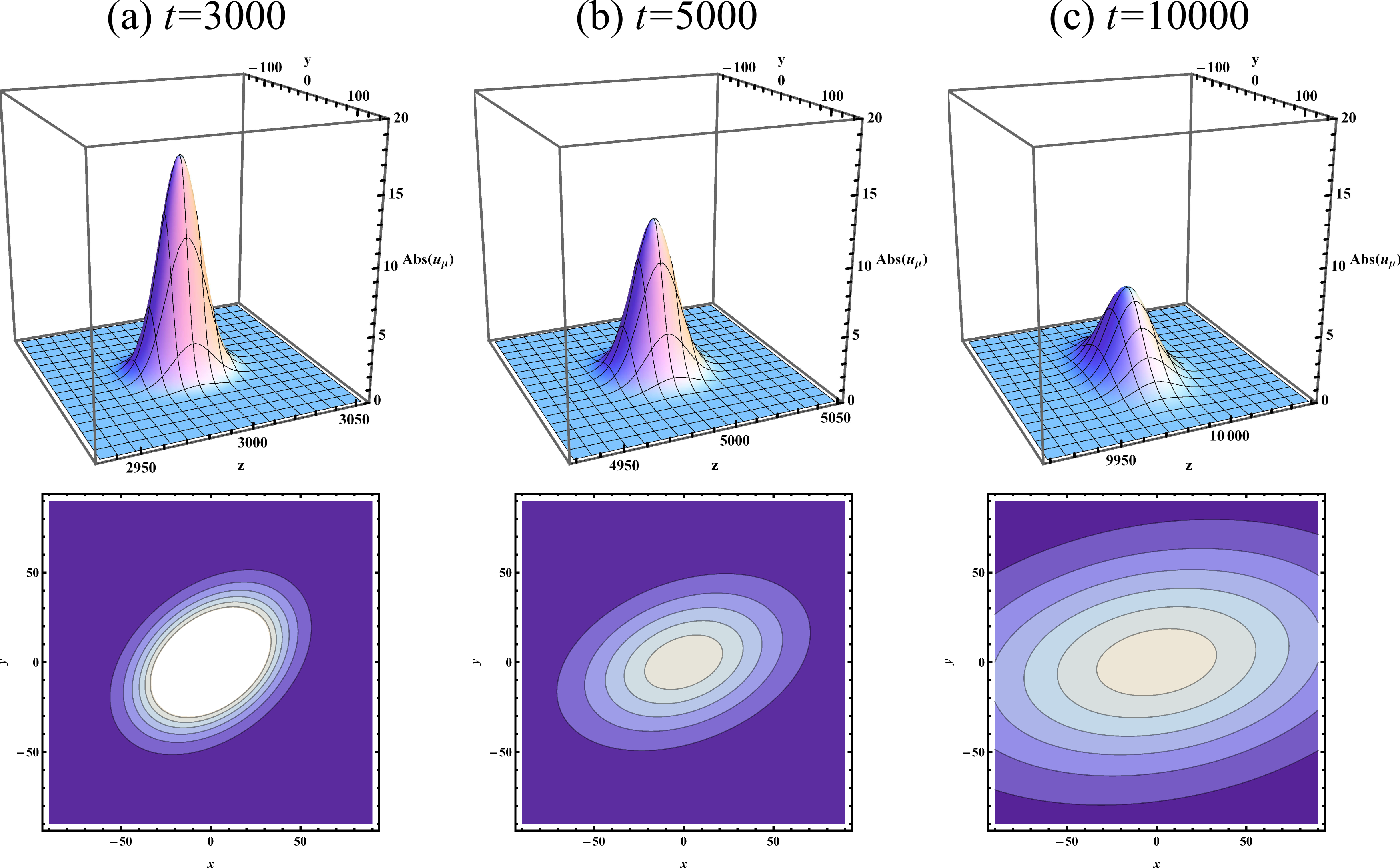}
\caption{Moderate times bahviour of absolute value of $u_p$ for consequent times $t=3000, 5000, 10000$ as a function of $y$ and $z$ for $x=0$ (upper pictures) and as function of $x$ and $y$ for $z=v_{gr}t$ (lower pictures). Convenient normalization for the $|u_p|$ is chosen, the coordinartes and time are in the units of mass $m$. See text for the values of all parameters.}
\label{moder_t_pic}
\end{figure}

Moderate times are characterized by condition
$ \|\G_0\|^{-1} \leq t \ll \tau$.  The solution in this
case can be described rather good by the asymptotic
formula \Ref{pack-mod-t} with $\zeta=2 t$. The distortion
of the solution in this regime is twofold. First of all,
the absolute value of the solution is decreasing linearly
with time   due to the dependence on $t$ of the prefactor
$A$ \Ref{pack-mod-t-A} via   $\sqrt{\det\G}$. Secondly,
the transverse width of the solution $\Delta_\perp$ also
grows 
linearly with time. Both these features are clearly
visible on the
Fig.\ref{moder_t_pic}, where the absolute value of
$u_p$ is plotted
at the same values of the parameters as before. We also
note that at
this stage the astigmatic properties can already be
seen --- the
localization ellipse is slowly rotating.

\begin{figure}
\centering
\includegraphics[width=5in]{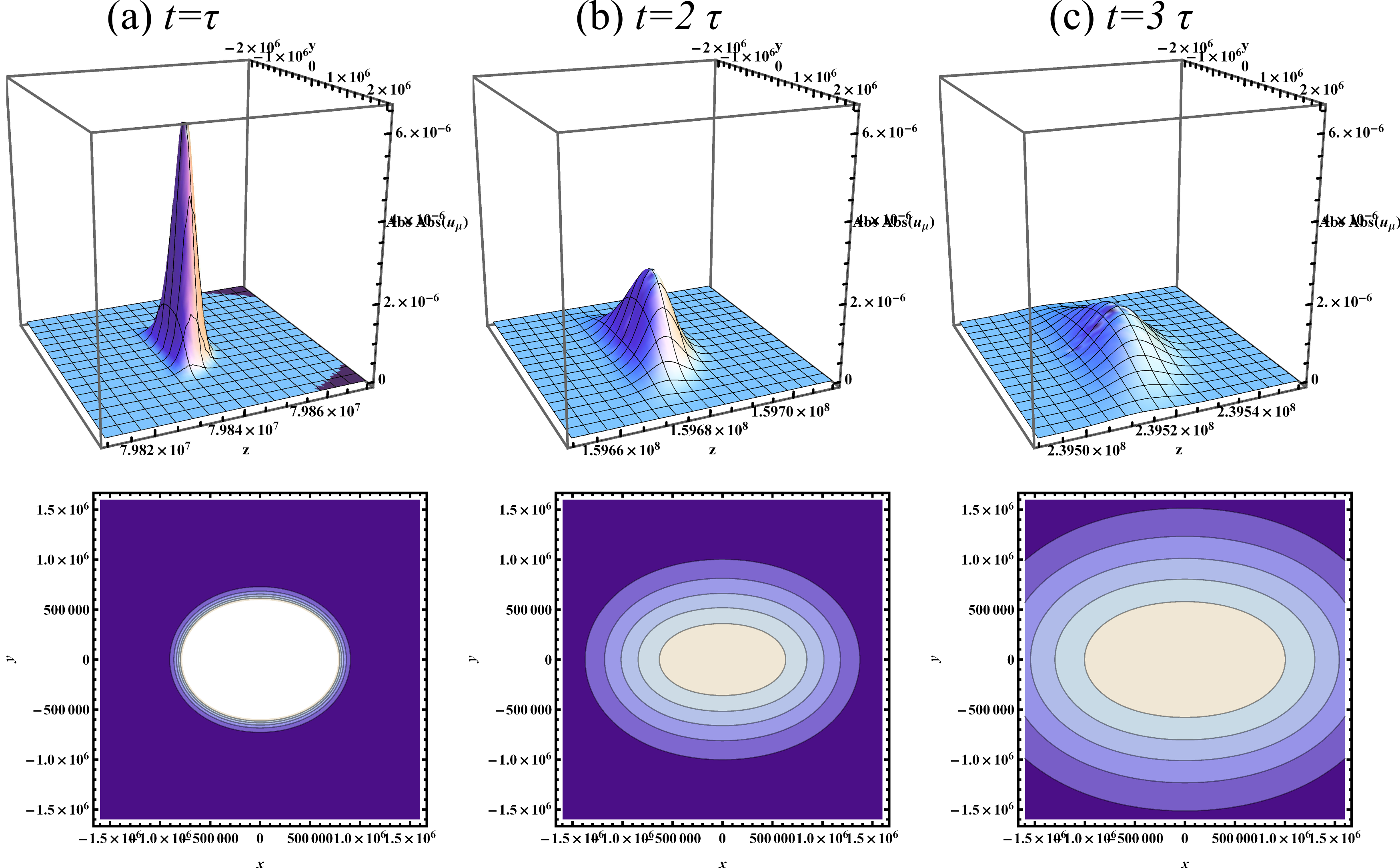}
\caption{Large  times bahviour of absolute value of $u_p$, for consequent times $t=\tau,2\tau,3 \tau$ as a function of $y$ and $z$ for $x=0$ (upper pictures) and as function of $x$ and $y$ for $z=v_{gr}t$ (lower pictures). Convenient normalization for the $|u_p|$ is chosen, the coordinartes and time are in the units of mass $m$. See text for the values of all parameters.}
\label{large_t_pic}
\end{figure}

For the large time regime $t$ satisfies conditions
(\ref{lar-t-defin}). The amplitude of the solution
also decreases  with time in this case, as it follows from
\Ref{A_large_t}.
The  packet  is concentrated in the intersection of a
cone and a spherical annulus as it is seen from
\Ref{lar-t-pack}. The width  of the annulus increases
linearly with time as $\Delta_v t$, while the angular
width of the cone is given by $\Delta_{\vartheta}$,
see  \Ref{lar-t-Del1}, \Ref{lar-t-Del2}. Their connection
with the widths at small times is
\be\label{lar-t-par-widths}
    \Delta_v^2 = p \Delta_{\parallel}^4, \quad
        \Delta_{\vartheta}^2 = \frac{1}{({\cal K}
        \Delta_{\perp})^2}.
 \ee
We see that the larger is the transverse width for
small time the narrower is a cone. This property
reflects the uncertainty principle.
 Large-time behaviour is presented on the
 Fig.\ref{large_t_pic}, where the absolute value of
 $u_p$ is plotted over quite large period of time at
 the same values of the parameters as before.
 The astigmatic properties are already frozen at
 this regime, the axis of the localisation ellipse are
 rotated to the full angle $\pi$ as compared with its
 position at $t\to-\infty$. The 
 latter fact can be understood  if one notes that the
 localization ellipse 
 for large times is defined by $\G_0^{-1}$,
 see \Ref{lar-t-Del2}, contrary to the case of small
 times where it is defined by $\G_0$. Under inversion,
 the smaller eigenvalue (the ellipse 
 axis) becomes the bigger one, so, the localization ellipse 
 effectively 
 rotates by an angle of $\pi/2$ as compared to small times.

From a practical point of view, it can be more
convenient to characterize the solutions not by
choosing the initial  parameters $\tau$, $\g$ and
$\e$, but by specifying their asymptotic properties ---
the wave number,  group velocity and localization widths
either at small times, or at large ones.
The former are unambiguously 
expressed through the latter
as we see from \Ref{lar-t-par-widths}.

 A legitimate practical problem is to find for a given
 KGFE (i.e. for a given value of $m$) a solution with
 particular values of, e.g.,  $\Omega$ and the width of
 the packet for small times $\Delta_{\parallel}$.
 From  \Ref{pack-mod-t-ds} we can deduce then the value
 of $p$, which must satisfy $p \gg 1$, if we wish the
 asymptotics be applicable 
 Next, we calculate the wave number,
 ${\cal K}^2 = \Omega^2 - p^2$ and the group speed,
 $v_{gr} = {\cal K}/\Omega$. Knowing $p$ and $v_{gr}$
 we can derive both the product $\tau\g$ by \Ref{par-p} and the
 ratio $\g/\tau = (1 - v_{gr})/(1 + v_{gr})$, which
 together gives us all the parameters of the
 desired 
 solution but its astigmatic properties. We can deduce
 the latter by choosing, for instance, the transversal
 width at small times, $\Delta_\perp$. Now all the
 parameters for the solution are known. Parameter
 $\mu$ cannot be derived by considering the asymptotic
 properties in the highest order.


We have however some restrictions to be satisfied,
if we wish our solution possesses good
localization 
properties
\be\label{p-K-cond}
    p \ll {\cal K}^2\frac{\Omega+{\cal K}}{\Omega-{\cal K}} , \quad p^3 \ll (\Omega {\cal K})^2,
    \quad p \ll {\cal K}^2 \frac{\varepsilon}{\tau}.
\ee
The first condition makes a longitudinal width of the
packet for small and moderate times smaller than the
distance where asymptotics works, see \Ref{mod-t-cond-t}.
The second and the third ones concern the large-time
asymptotics, they originate from \Ref{cond-loc}. The
second condition ensures that the speed of increasing
of the longitudinal width of the packet is smaller
then the group speed. The third condition means that
the angle of  the cone is small. All of these conditions
can be satisfied, for example, if we assume that
$\Omega \sim p$, $\Delta_{\parallel} \sim 1/\sqrt{p}$.
Then  $\cal{K}$ is of order of $O(p)$. If we take
parameters in such a way that ${\cal K} \gg \sqrt{p}$
as well, the restriction conditions \Ref{p-K-cond} will
be satisfied.

We expect that our results may be useful for prediction
of waves propagation in media with dispersion. The
obtained solution have non-zero angular momentum \cite{Angular}
which is studied intensively for the wave equation
in context of manipulating 
of nanoparticles.
 The investigation of  this momentum for the waves
 in dispersive media is a very appealing for the 
 future research. Another possible application
 concerns two-dimensional solutions of KGFE. Such solution
 may be a base for further design of localized 
 solutions of the Dirac equations which may find
 application in prediction and modelling
 of waves in epitaxial graphen.


\vskip.5cm
\section*{Acknowledgements}
This work was supported in part by  FAPESP (I.V.F.)
and RFBR grant  140200624 (M.V.P.).


\appendix

\section{Asymptotic behaviour of KGF
solutions }\label{Asymptotic-KGF}

\subsection{General properties of solutions of the KGF
equation}
We show here how the properties of solutions of the KGFE
can be found within a general approach based on Fourier
representation.

Any solution of the KGF equation can  be written as a
Fourier
integral
\begin{equation}\label{as-t1}
u(\br,t) = \frac{1}{(2\pi)^{n+1}}
    \int\limits_{\mathbb{R}^{n}} d^n\bk d\om \,\,
     \hat{u}(\o,\bk ) \exp{i \bk\br - i\omega t }
     \delta(\om-\om(\bk)),
\end{equation}
where $\omega(\bk )=\sqrt{\bk^2+m^2}$ , and $n=d+1$.
It is convenient to introduce new
dimensionless variables
$\bnk=\bk/m,$ $\no=\o/m$.  We assume that the
modulus of the
Fourier transform $\hat{u}(\bk )$ has a sharp maximum
and $\hat{u}(\bk )$ can be written in the form
\be\label{Four-int}
 \hat u(\bk) = a(\bnk) \exp{-p \Phi( \bnk)}, \quad p \gg 1,
\ee
i.e. we assume that $\Phi$ has stationary point in the
minimum of its real
part. The formula (\ref{as-t1}) may be rewritten in the
form suitable for the analysis
by the method of the steepest descent now
\be\label{int-steep}
u(\br,t) = 
\frac{m^n}{(2\pi )^{n+1}}
    \int\limits_{\mathbb{R}^{n}} d^n {\bm \chi} \,\,
     a( {\bm \chi} ) \exp{- p \Psi( {\bm \chi})},
\ee
where
\be\label{phase-int-steep}
\Psi( \bnk) = \Phi( \bnk) + i \tilde{t} (\no  -  \bnk\bv),
\quad
    \bv=\br/t, \quad
    \tilde{t} = m t /p, \quad
    \no = \sqrt{\nk^2 + 1}.
\ee
The main term of asymptotics of the integral
(\ref{int-steep}) for $p \to \infty$ reads \cite{}
\begin{equation}\label{as-t22}
  u(\br,t) \mathop{\simeq}_{p\to\infty}
    \exp{-p \Psi( \bnk_*)} a( \bnk _*) \frac{m^n}{(2\pi)^n}
    \sqrt{\frac{(2 \pi)^n}{p^n \det \Psi''_*} }
      ( 1 + O(p^{-1})),
\end{equation}
where   $\det \Psi''_*$ is the
determinant of the Jakobi matrix $ \Psi''$
calculated in the saddle
point $\bnk_*$, i.e., the matrix of the second
derivatives of $\Psi$ with respect to $\bnk$.
The saddle point
$\bnk_*$ should be found from the equation
\be\label{saddle-p-mod-t}
\nabla \Psi(\bnk_*) =
    \nabla \Phi(\bnk_*) + i \tilde{t}(\nabla
    \no(\bnk_*) - \bv) = 0.
\ee
For  small times (in
comparison with $p/m$)  we seek the saddle point as an
expansion
$\bnk_* = \bnk_*^{(0)} + i \tilde{t}  \bnk_*^{(1)}
+ \ldots$
and get
\be
\nabla \Phi(\bnk_*^{(0)}) = 0, \quad \bnk_*^{(1)} =
    -  (\Phi''_0 )^{-1}(\bv_{gr} -\bv), \quad \bv_{gr}
    \equiv
    \nabla \no(\bnk_*^{(0)}),
\ee
where $\Phi''_0$ is the matrix of
the second
derivatives of $\Phi$ with respect to $\bnk$ calculated
in the point
$\bnk_*^{(0)}$. Corrections   are to be taken into account
in the formula (\ref{as-t22}) only in the exponential term
containing the large parameter $p$
\be\label{exp-Psi}
 \Psi(\bnk_*) \simeq  \Phi(\bnk^{(0)}_*) + i  \tilde{t}
    (\no^{(0)}_*  - \bnk^{(0)}_*\bv) -
    \frac{ \tilde{t}^2}{2}
     \( \Phi''_0  \bnk_*^{(1)}, \bnk_*^{(1)}\) -
    \tilde{t}^2 \((\bv_{gr} - \bv), \bnk_*^{(1)} \).
\ee
Substituting (\ref{exp-Psi}) in the (\ref{as-t22}),
neglecting the
correction terms in the amplitude and recalling that
$\bv t = \br$ we obtain
\begin{eqnarray}\label{as-t22-fin}
&&  u(\br,t) \approx {\cal A}_1 \,
\exp{ - i (\omega_0 t - \bk_0 \cdot \br)}
\exp{ - \frac{m^2}{2p} \( (\br - \bv_{gr}t),
(\Phi''_0)^{-1} (\br - \bv_{gr}t) \)},\\
&&  {\cal A}_1 = \frac{m^n }{2\pi (2 \pi p)^{n/2} }
\frac{\hat{u}(\bk_0)}{ \sqrt{\det \Phi''_0}}
, \quad \bk_0 = m \nk_*^{(0)},
\quad \omega_0 = m
\no(\nk_*^{(0)}).
\end{eqnarray}
This formula  can be applied to the
exact solution $u_p$ presented in Section
\ref{loc-part-KGF}.  Its  Fourier image
(\ref{u_p(k)}) can be given in the form
(\ref{Four-int}) as follows
\begin{eqnarray}
&& a(\bnk) =  \frac{\hat{C}_p}{m^{\nu+1+n/2}}\,\,
        \frac{1}{ \no(\no+\nk_z)^{\nu +n/2}}, \label{amp-int} \\
&& \Phi(\bnk) =  i \tfrac{(\nk_\perp,\G^{-1}_0\nk_\perp)}
        {2\sqrt{\g\tau}(\no+\nk_z)}
    +\tfrac{\sqrt{\g/\tau}}{2} (\no+\nk_z)
    + \tfrac{\sqrt{\tau/\g}}{2(\no+\nk_z)}
        +,\quad
    p = m \sqrt{\tau \g}.
\label{phase-int}
\end{eqnarray}
It is easy to check that formula (\ref{as-t22-fin})
with account of (\ref{amp-int}) and (\ref{phase-int})
gives (\ref{pack-mod-t}).


\subsection{Large-time behaviour of the particle-like
solution}

Now we turn to the large time behaviour and
assume that $\hat{u}(\bk)$ in (\ref{as-t1})  changes
slowly as
compared  with the oscillatory term. Thus we are able
to proceed
with stationary phase method and obtain \cite{Whitham74}
\begin{equation}\label{as-1-t}
u(\br,t) \mathop{\simeq} \frac{\hat{u}(\bk _*)}
{(2\pi)^{n/2}} \,\, \frac{1}{|\det (\o''_* t)|^{1/2}}
    \exp{ - i t ( \o(\bk_*)  - \bk _*\bv) -
    i\frac{n\pi}{4}{\rm sgn}(t)}.
\end{equation}
Here $\bk _*=\bk_*(\bv)$ is the solution of the equation
\begin{equation}
  \nabla \omega(\bk _*) =
  \bv \equiv {\br}/{t},\label{gr-sp}
\end{equation}
where $\o(\bk_*) = \sqrt{\bk_*^2 + m^2}$. It is easy
to check that $\bk_*=m\bnk$, $\o(\bk_*)=m\no$ where
$\bnk(\bv)$ and $\no(\bv)$ are given by (\ref{no-nk}).
By $\omega''_*$ in (\ref{as-1-t}) we denote the
$n\times n$ matrix of second derivatives of $\o$
with respect to components of $\bk$ calculated in
the point $\bk _*(\bv)$. It is easy to check that
\be
\det\omega''_*=\frac{1}{\omega^{n}_*} -
    \frac{\bk^2_*}{\omega^{n+2}_*}=
    \frac{m^2}{\omega^{n+2}_*}
            = \frac{(1-v^2)^{n/2+1}}{m^{n} }\,.
    \label{parav-l-t1}
\ee
Formula (\ref{as-1-t}) demonstrates complicated dependence
of $\br$
and $t$ through $\bv$. It reads
\begin{equation}\label{as-l-t3}
u(\br,t)
    \mathop{\simeq}_{t\to\infty}
    \frac{m^{n/2}}{(2 \pi)^{n/2}}
       \frac{\exp{ - i m\sqrt{t^2-r^2}}  }
       {\, |t|^{n/2}\,(1-v^2)^{(n+2)/4}}
    e^{-i \frac{n\pi}{4}{\rm sign}(t)}
    \, \hat{u}\(\frac{m \bv}{\sqrt{1 - v^2}}\)\,.
\end{equation}

After substitution of $\hat{u}$ from (\ref{u_p(k)})
with account of (\ref{no-nk}) we obtain the formula
which is in agreement with previously found formula
(\ref{lar-t-pack}).
It is important to note that the obtained formula
cannot be applied
when $m \to 0$. It is due to the fact that the second
derivative of the phase function from  (\ref{as-1-t})
in this case tends to zero, thus the region of validity
of the asymptotic (\ref{as-l-t3}) is approaching spacial
infinity.  The stationary phase method which we used is
not applicable in this case.




\end{document}